\documentclass[preprint,showpacs,superscriptaddress,amsmath,amssymb]{revtex4}
\usepackage{graphicx}
\usepackage{bm}

\begin{document}

\title{Volatility return intervals analysis of the Japanese market}

\author{Woo-Sung Jung}
\email{wsjung@physics.bu.edu}
\affiliation{Center for Polymer Studies and Department of Physics, Boston University, Boston, MA 02215, USA}
\affiliation{Center for Complex Systems and Department of Physics, Korea Advanced Institute of Science and Technology, Daejeon 305-701, Republic of Korea}
\author{Fengzhong Wang}
\affiliation{Center for Polymer Studies and Department of Physics, Boston University, Boston, MA 02215, USA}
\author{Shlomo Havlin}
\affiliation{Center for Polymer Studies and Department of Physics, Boston University, Boston, MA 02215, USA}
\affiliation{Minerva Center and Department of Physics, Bar-Ilan University, Ramat-Gan 52900, Israel}
\author{Taisei Kaizoji}
\affiliation{Division of Social Sciences, International Christian University, Tokyo 181-8585, Japan}
\author{Hie-Tae Moon}
\affiliation{Center for Complex Systems and Department of Physics, Korea Advanced Institute of Science and Technology, Daejeon 305-701, Republic of Korea}
\author{H. Eugene Stanley}
\affiliation{Center for Polymer Studies and Department of Physics, Boston University, Boston, MA 02215, USA}

\date{\today}

\begin{abstract}
We investigate scaling and memory effects in return intervals between price volatilities above a certain threshold $q$ for the Japanese stock market using daily and intraday data sets. We find that the distribution of return intervals can be approximated by a scaling function that depends only on the ratio between the return interval $\tau$ and its mean $\left<\tau\right>$. We also find memory effects such that a large (or small) return interval follows a large (or small) interval by investigating the conditional distribution and mean return interval. The results are similar to previous studies of other markets and indicate that similar statistical features appear in different financial markets. We also compare our results between the period before and after the big crash at the end of 1989. We find that scaling and memory effects of the return intervals show similar features although the statistical properties of the returns are different.
\end{abstract}


\maketitle

\section{Introduction}
In recent years, financial markets have been studied using statistical physics approaches \cite{mantegna00,mantegna1995,johnson03,scalas00,takayasu97,tsallis03,lillo00} and some stylized facts have been observed including (i) the probability density function (pdf) of the logarithmic stock price changes (log-returns) has a power-law tail \cite{mandelbrot1963,pagan96,gopikrishnan99,liu99,gabaix03}, (ii)  the absolute value of log-returns are long-term power-law correlated \cite{pagan96,ding83,ord85,harris86,schwert89,granger96,plerou01,plerou05}. Statistical properties of price fluctuations are important to understand market dynamics, and are related to practical applications \cite{bouchaud03}. In particular, the volatility of stocks attracted much attention because it is a key input of option pricing models such as the Black-Scholes \cite{black73,engle82,cox76,campbell97}.

Yamasaki \textit{et al.} \cite{yamasaki05} investigated the return intervals between volatility above a certain threshold in the US stock and foreign exchange markets. They analyzed \textit{daily} data and found scaling and memory effects in return intervals. Wang \textit{et al.} \cite{wang06} studied the return intervals in \textit{intraday} data of the US market, and found similar scaling and memory effects. Weber \textit{et al.} \cite{weber07} analyzed the memory in volatility return intervals. In this manuscript, we further test the generality of the above findings in the Japanese stock market data where we include both \textit{daily} and \textit{intraday} data sets. We find that also in this case the pdf of return intervals mainly depends on the scaled parameter; the ratio between the return intervals and their mean. Memory effects also exist in the return intervals sequences.

In addition, we study scaling and memory effects considering different type of market dynamic periods. The Japanese market in recent decades (1977-2004) can be divied into two periods, the inflationary (before 1989) and the deflationary (after 1989). Kaizoji \cite{kaizoji04b} showed that some statistical properties of the returns are different in the two periods. The absolute return distribution in the inflationary period behave as a \textit{power-law} distribution, while the return distribution in the deflationary period obeys an \textit{exponential law}. Here, we find that scaling and memory effects of the return intervals show similar features in both periods.

\section{Scaling and memory properties}
In this section, we analyze the statistical properties of the Japanese stock market using \textit{daily} and \textit{intraday} return intervals. We investigate the \textit{daily} data of three representative companies, Nippon Steel, Sony and Toyota Motor listed on the Tokyo Stock Exchange (TSE) for the 28-year period from 1977 to 2004, a total of 7288 trading days. Also, we study the \textit{intraday} data of 1817 companies listed on the TSE from January 1997 to December 1997. The sampling time is 1 minute and the data size is about 9 million.

The logarithmic return $G(t)$ is written as $G(t)\equiv \ln Y(t+\Delta t)-\ln Y(t)$ where $Y(t)$ is the stock price at time $t$, and the normalized volatility $g(t)$ is defined as:
\begin{equation}
g(t)\equiv\frac{|G(t)|}{\sqrt{\left<G(t)^2\right>-\left<G(t)\right>^2}},
\end{equation}
where $\left<\cdots\right>$ means time average. We pick every \textit{event} of volatility $g(t)$ above a certain thresold $q$. The series of the time intervals between those events, depending on the threshold $q$, $\{ \tau (q) \}$, are generated.

\begin{figure}
\includegraphics[width=1.0\textwidth]{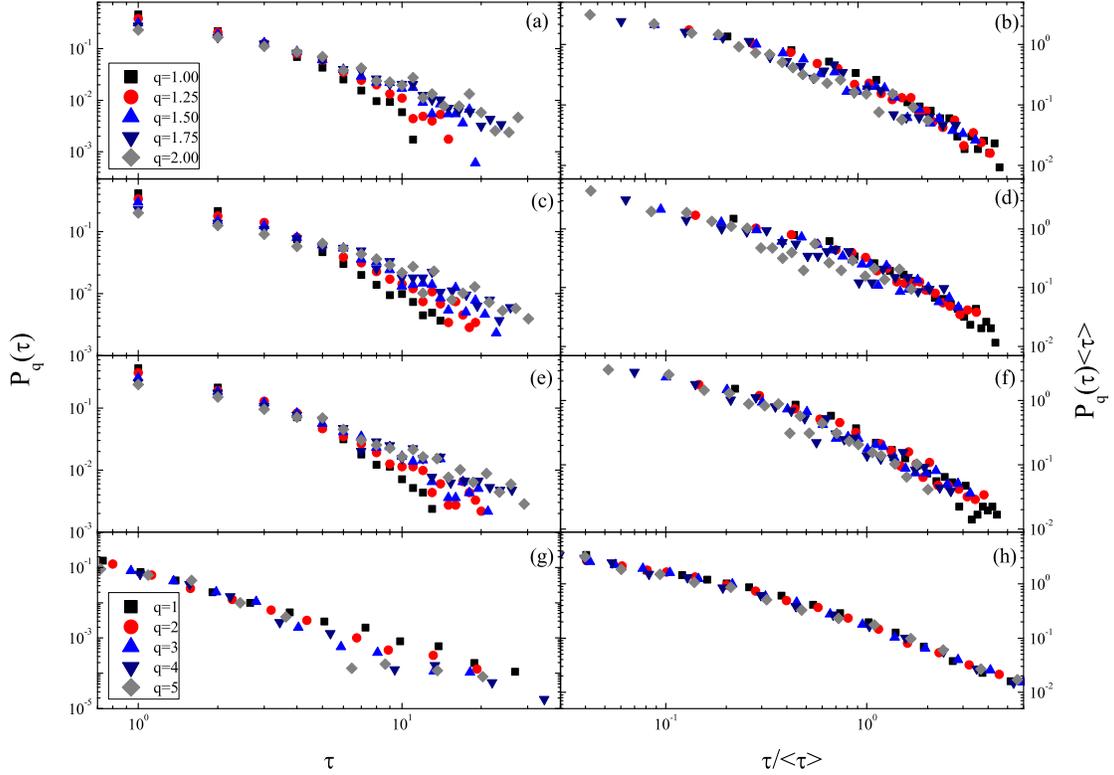}
\caption{\label{fig:pdf_daily} (Color online) Distribution and scaling of return intervals using for (a) and (b) Nippon Steel, (c) and (d) Sony, (e) and (f) Toyota Motor, and (g) and (h) mixture of 1817 Japanese companies. Daily data is used for (a)-(f), and intraday for (g) and (h). The sampling time for intraday data is 1 min. Symbols represent different threshold $q$ varying from 1 to 2 (for (a)-(f)) and 1 to 5 (for (g) and (h)), respectively.}
\end{figure}

We investigate the pdf $P_q(\tau)$ to better understand its behavior and how it depends on the threshold $\tau$ (the left panels of Fig. \ref{fig:pdf_daily}). The scaled pdf, $P_q(\tau)\left<\tau\right>$, as a function of the scaled return intervals $\tau/\left<\tau\right>$ is shown in the right panels of Fig. \ref{fig:pdf_daily}. Previous study \cite{kaizoji04} showed the distributions of $P_q(\tau)$ are different with different threshold $q$, and we find the same result. However, when plotting $P_q(\tau)\left<\tau\right>$ as a function of $\tau/\left<\tau\right>$, we obtain an approximate collapse onto a single curve. The collapse means that the distributions can be well approximated by the scaling relation

\begin{equation}
P_q(\tau)=\frac{1}{\left<\tau\right>}f(\frac{\tau}{\left<\tau\right>}).
\label{eq:p}
\end{equation}

The scaling function $f(\tau/\left<\tau\right>)$ of Eq. (\ref{eq:p}) does not depend directly on the threshold $q$ but only through $\left<\tau\right>\equiv\left<\tau(q)\right>$. Therefore, if $P_q(\tau)$ is known for one value of $q$, the distribution for other $q$ values can be predicted using Eq. (\ref{eq:p}). Figs. \ref{fig:pdf_daily}g and \ref{fig:pdf_daily}h show that the same features, distribution and scaling of return intervals exist for \textit{intraday} data after removing the intraday trends \cite{dacorogna01}. The size of the intraday data set is basically larger than that of daily data, and consists of 1817 companies. Therefore, we are able to extend our study to larger values of $q$ and get better statistics (less scattering) compared to those in Fig. \ref{fig:pdf_daily} (a)-(f).

Previous similar studies on the US stock and foreign exchange market \cite{yamasaki05,wang06} suggested that there might be a universal scaling function for the return time intervals of different financial markets. We observe the same result also for both daily and intraday data of the Japanese market, and it raises the possibility that the scaling function is universal.

\begin{figure}
\includegraphics[width=1.0\textwidth]{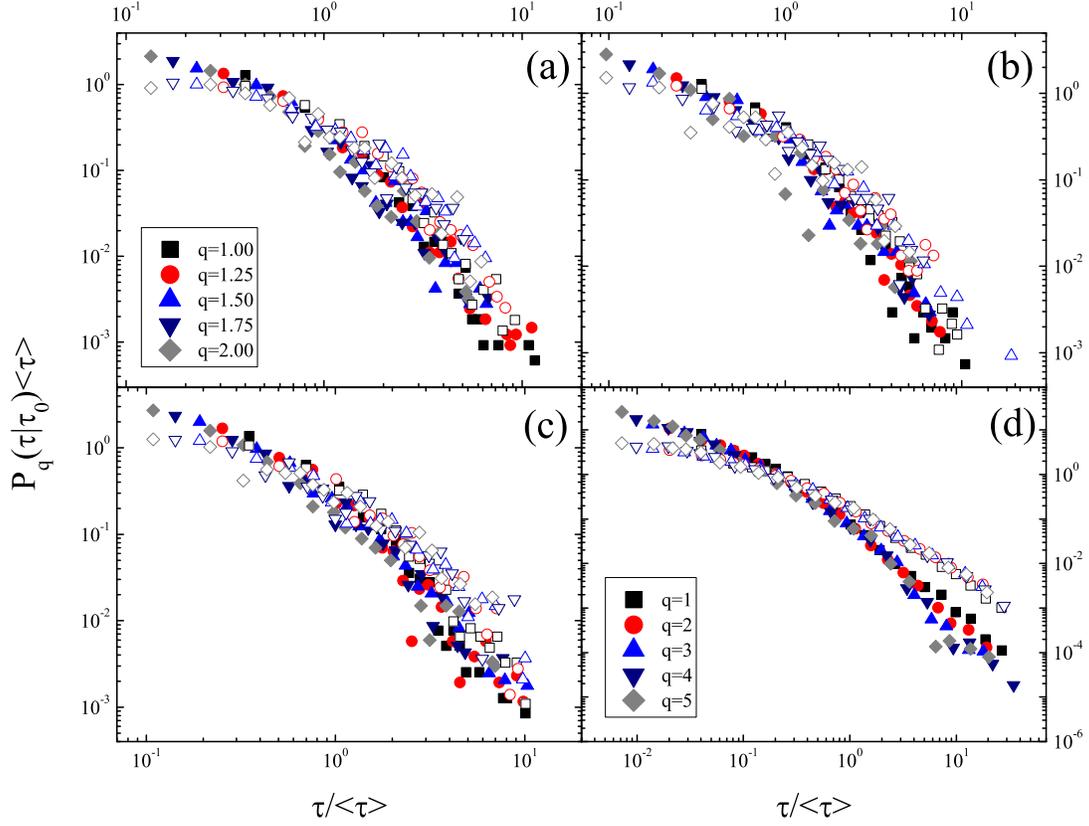}
\caption{\label{fig:memory_daily} (Color online) Scaled conditional distribution $P_q(\tau|\tau_0)\left<\tau\right>$ as a funtion of $\tau/\left<\tau\right>$ using daily data for (a) Nippon Steel, (b) Sony, (c) Toyota Motor, and (d) mixture of 1817 companies. Symbols represent different threshold $q$.}
\end{figure}

We also test whether the sequence of the return intervals is fully characterized by the distribution $P_q(\tau)$. If the sequence of $\tau$ are uncorrelated, the return intervals are independent of each other and chosen from the probability distribution $P_q(\tau)$. However if they are correlated, the memory also affects the order in the $\{\tau\}$ time sequence. $P_q(\tau|\tau_0)$ represents the conditional pdf which is the probability of finding a return interval $\tau$ following a return interval $\tau_0$. If memory does not exist, we expect that the conditional pdf will be independent of $\tau_0$ and identical to $P_q(\tau)$. We study $P_q(\tau|\tau_0)$ for a range of $\tau_0$ values. The full data set of $\{\tau\}$ is divided into eight subsets with return intervals in increasing order. We show $P(\tau|\tau_0)$ for $\tau_0$ being in the lowest subset (full symbols) and, in the largest subset (open symbols) in Fig. \ref{fig:memory_daily}. The results show that for $\tau_0$ in the lowest subset, the probability of finding $\tau$ below $\left<\tau\right>$ is enhanced compared to $P_q(\tau)$, while the opposite occurs for $\tau_0$ in the largest subset. The pdfs, $P_q(\tau|\tau_0)$, for all thresholds collapse onto a single scaling function for each $\tau_0$. This suggests that $P_q(\tau)$ does not characterize the sequence of $\tau$ and memory exists in the sequence.

\begin{figure}
\includegraphics[width=1.0\textwidth]{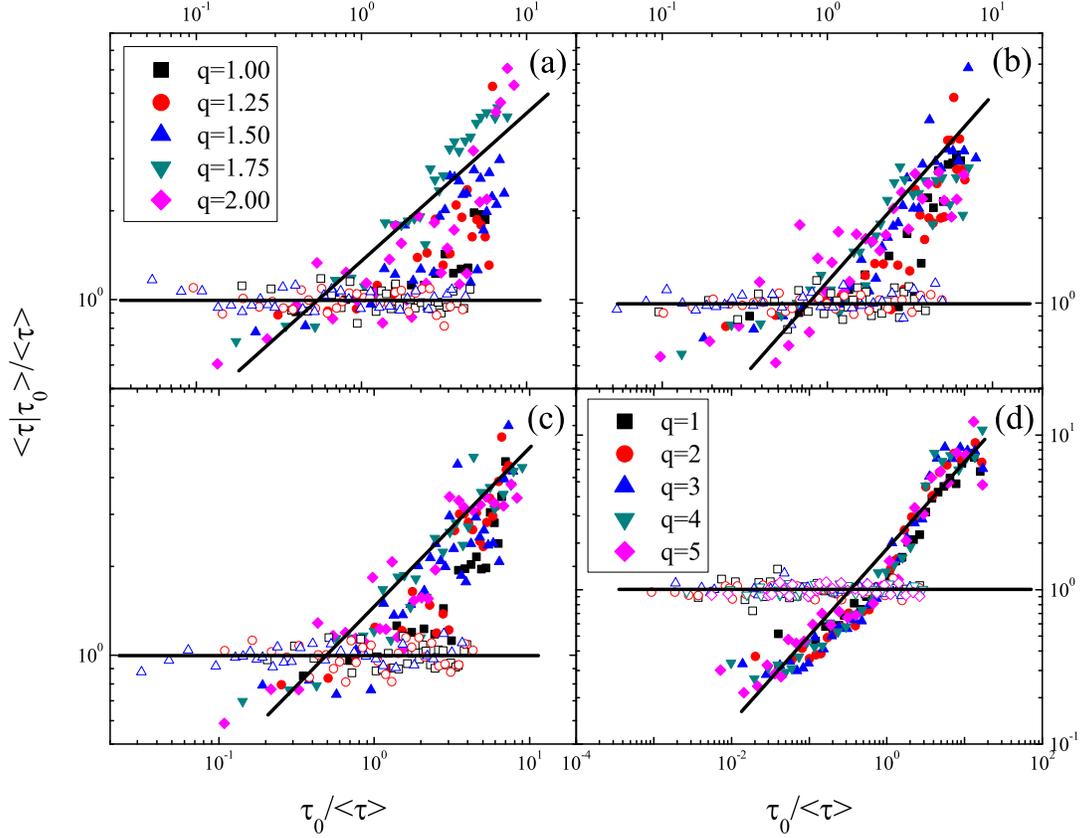}
\caption{\label{fig:mean} (Color online) Scaled mean conditional return interval $\left<\tau|\tau_0\right>/\left<\tau\right>$ as a function of $\tau_0/\left<\tau\right>$ for (a) Nippon Steel, (b) Sony, (c) Toyota Motor, and (d) mixture of 1817 companies. Open symbols correspond to shuffled data. The lines serve only as a guide to the eyes.}
\end{figure}

The memory effects are also observed in the mean conditional return interval $\left<\tau|\tau_0\right>$ which is the first moment of $P_q\left<\tau|\tau_0\right>$  shown in Fig. \ref{fig:mean}, where we plot $\left<\tau|\tau_0\right>/\left<\tau\right>$ as a function of $\tau_0/\left<\tau\right>$. It is seen clearly that large (or small) $\tau$ tend to follow large (or small) $\tau_0$. Note that the shuffled data (open symbols) exhibit a flat shape, which means $\tau$ is independent on $\tau_0$. The above results show that the return intervals $\tau$ strongly depend on the previous return interval $\tau_0$.

\begin{figure}
\includegraphics[width=1.0\textwidth]{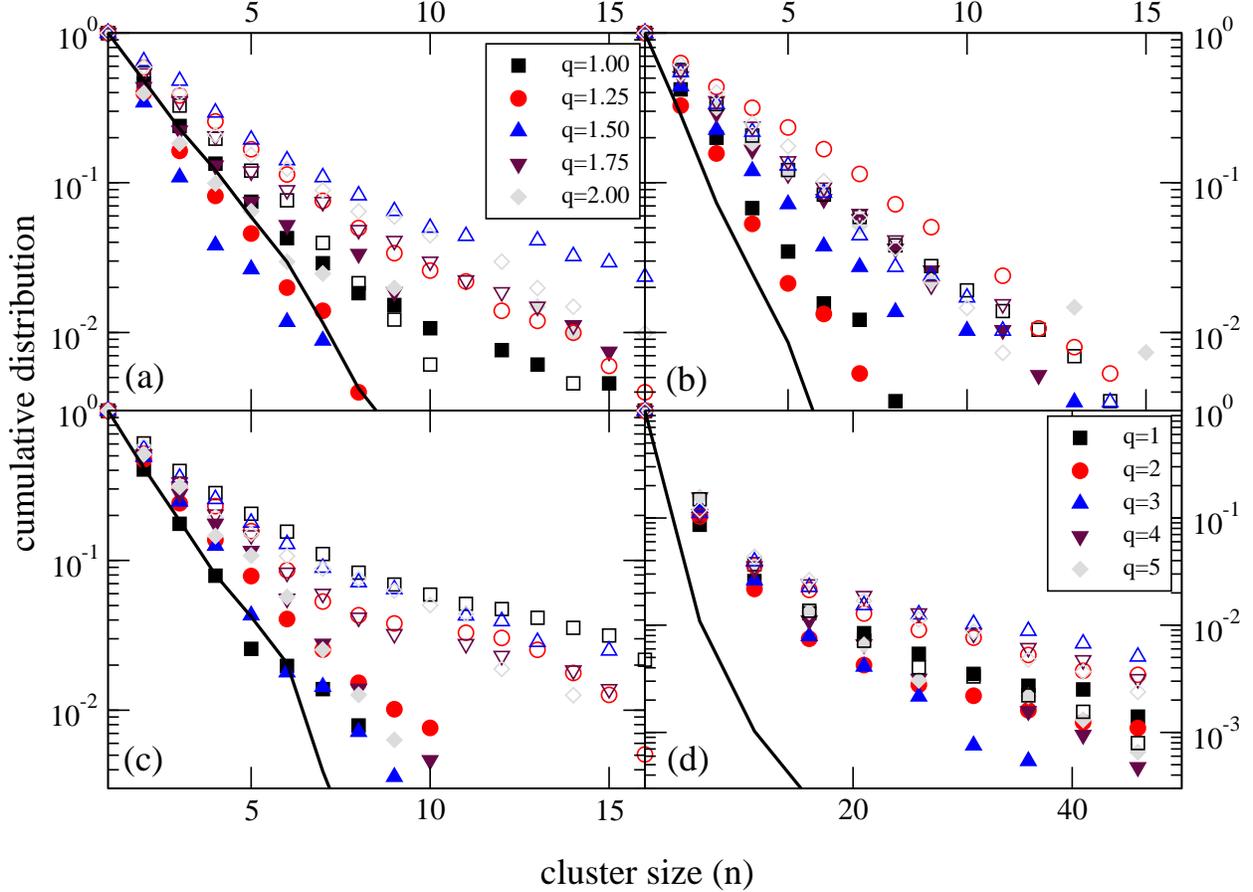}
\caption{\label{fig:cum_histo} (Color online) Cumulative distribution of size for return intervals clusters for (a) Nippon Steel, (b) Sony, (c) Toyota Motor, and (d) mixture of 1817 companies. The distributions consist of consecutive return intervals that are all above (closed symbols) or below (open symbols) the median of all the interval records. The straight lines show the shuffled volatility case (q=1, above the median) where memory is removed.}
\end{figure}

We also analyze clusters of short and long return intervals in order to investigate clustering phenomena, which represent further and longer term correlations compared to $P_q(\tau|\tau_0)$ and $\left<\tau|\tau_0\right>$. The sequence of return intervals is devided into two bins by the median of the entire database. The two bins consist of  intervals which are ``above" and ``below" the median respectively. A cluster is formed by $n$ consecutive return intervals that are ``above" or ``below" the median. Fig. \ref{fig:cum_histo} shows the cumulative distribution of clusters of size $n$ for three Japanese companies and mixture of 1817 companies. Both ``above" and ``below" clusters have long tails compared to the surrogate volatility shuffled case.

\section{Inflation and deflation}
\begin{figure}
\includegraphics[width=1.0\textwidth]{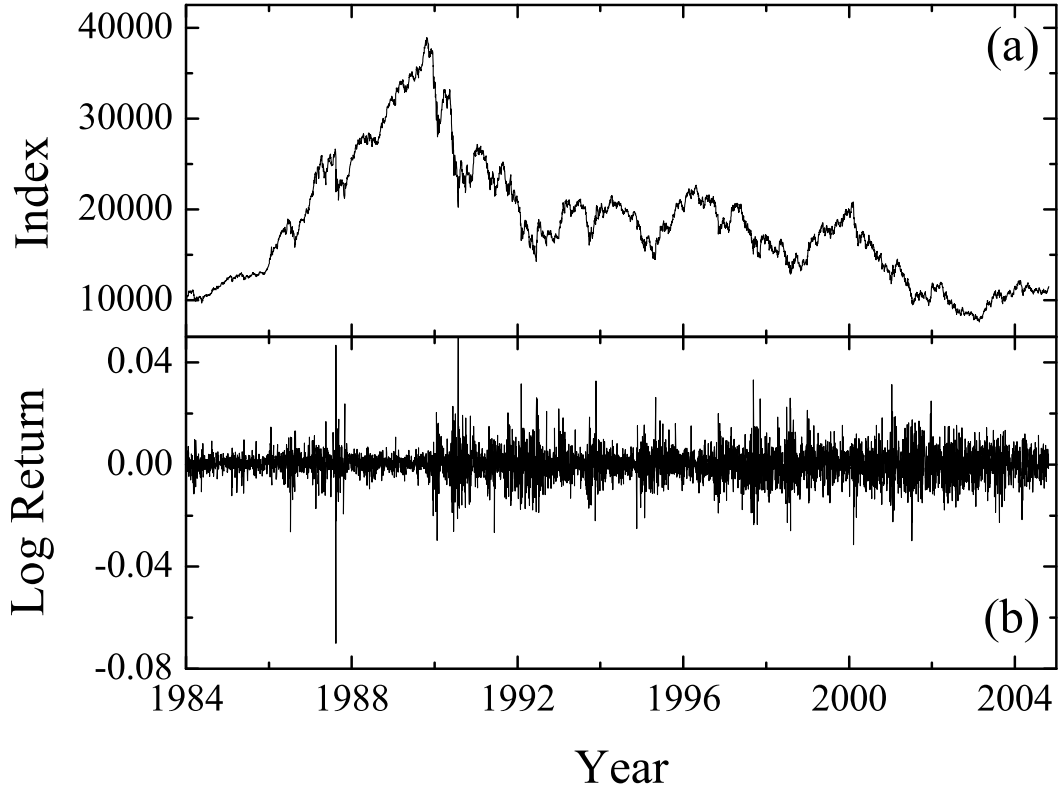}
\caption{\label{fig:nikkei} The time series of NIKKEI 225 (a) index and (b) log return from January 1984 to December 2004.}
\end{figure}

In this section, we investigate the NIKKEI 225 index data to answer one question: Even the return distributions have different features, do the return time intervals show similar features? The NIKKEI 225 index reached the highest position on the last trading day of 1989, but declined from the first trading day of 1990. It has dropped about 63 percent from 1990 to August of 1992. This is a famous Japanese market bubble and crash. Therefore, the Japanese market between 1977 and 2004 can be divided into two parts: the period of inflation, before December 1989, and the period of deflation, after January of 1990 (Fig. \ref{fig:nikkei}). Kaizoji \cite{kaizoji04b} showed that the return statistics of those two periods are clearly different. The return distribution in the inflationary period is approximated by an asymptotic power law, while the return distribution in the deflationary period seem to obey an exponential law.

\begin{figure}
\includegraphics[width=1.0\textwidth]{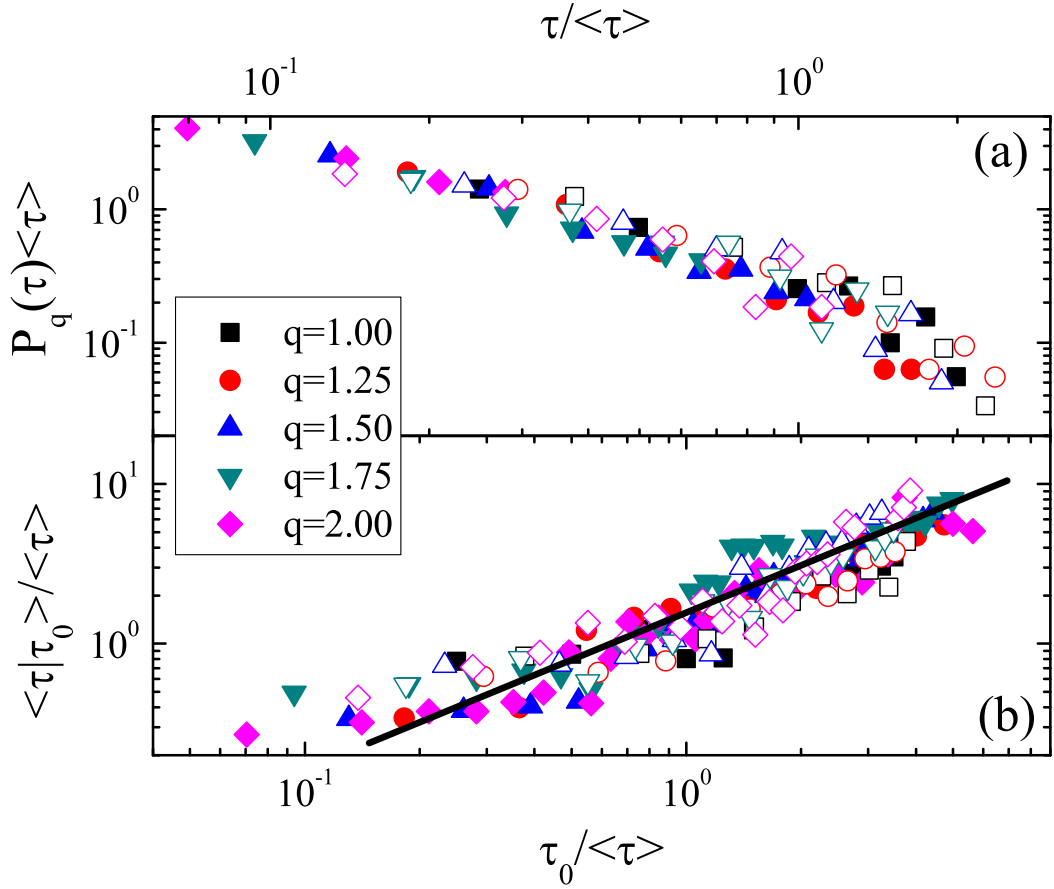}
\caption{\label{fig:inde} (Color online) (a) Distribution and scaling of return intervals and (b) scaled mean conditional return interval $\left<\tau|\tau_0\right>/\left<\tau\right>$ as a function of $\tau_0/\left<\tau\right>$. Full symbols correspond to the inflationary period and open ones the deflationary period. Symbols represent different threshold $q$. The line is only a guide to the eyes.}
\end{figure}

Fig. \ref{fig:inde}(a) represents the scaled pdf, $P_q(\tau)\left<\tau\right>$, as a function of the scaled return intervals $\tau/\left<\tau\right>$ in the inflationary period (full symbols) and the deflationary period (open symbols). No significant differences is seen between these two periods. Also, conditional mean return intervals of two periods show that $\tau$ depends on $\tau_0$ in a similar way in the two periods (Fig. \ref{fig:inde}(b)). It has been suggested that the pdf $P_q(\tau)$ and the scaled pdf $P_q(\tau)\left<\tau\right>$ are universal functions for different financial markets. Here we observe that even though the return distributions of the periods are different, the return intervals show similar features.

\section{Conclusions}
We investigated scaling and memory effects in volatility return intervals for the Japanese stock market using daily and intraday data sets. For both data sets, we found that the distribution of return intervals are well approximated by a single scaling function that depends only on the ratio $\tau/\left<\tau\right>$, and the scaling function is different from the Poisson distribution expected for uncorrelated records. Also, our results for the conditional distribution and mean return interval support the memory between subsequent return intervals, such that a large (or small) return interval is more likely to be followed by a large (or small) interval. The clustering shown in Fig. \ref{fig:cum_histo} shows that the memory exists even between nonsubsequent return intervals. Our results also support the possibility that the scaling and memory properties are similar functions for different financial markets. In addition, we tested scaling and memory effects in the inflationary and deflationary periods of the Japanese market. While the return distributions show different features, the scaling and memory properties of the return intervals are similar. It should be noted that similar scaling properties and memory in the return intervals have been found earlier also in climate \cite{corral04,bunde2004,bunde2005} and earthquakes \cite{livina2005}.



\begin{thebibliography}{00}
\bibitem{mantegna00}
R. N. Mantegna, and H. E. Stanley, \textit{An Introduction to Econophysics: Correlations and Complexity in Finance} (Cambridge University Press, Cambridge, 2000).

\bibitem{mantegna1995}
R. N. Mantegna, and H. E. Stanley, Nature \textbf{376}, 46 (1995).

\bibitem{johnson03}
N. F. Johnson, P. Jefferies, and P. M. Hui, \textit{Financial Market Complexity} (Oxford University Press, Oxford, 2003).

\bibitem{scalas00}
E. Scalas, R. Gorenflo, and F. Mainardi, Physica A \textbf{314}, 749 (2002).

\bibitem{takayasu97}
H. Takayasu, A. H. Sato, and M. Takayasu, Phys. Rev. Lett. \textbf{79}, 966 (1997).

\bibitem{tsallis03}
C. Tsallis, C. Anteneodo, L. Borland, and R. Osorio, Physica A \textbf{324}, 89 (2003).

\bibitem{lillo00}
F. Lillo, and R. N. Mantegna, Phys. Rev. E \textbf{62}, 6126 (2000).

\bibitem{mandelbrot1963}
B. B. Mandelbrot, J. Business \textbf{36}, 394 (1963).

\bibitem{pagan96}
A. Pagan, J. Empir. Finance \textbf{3}, 15 (1996).

\bibitem{gopikrishnan99}
P. Gopikrishnan, V. Plerou, L. A. N. Amaral, M. Meyer, and H. E. Stanley, Phys. Rev. E \textbf{60}, 5305 (1999).

\bibitem{liu99}
Y. Liu, P. Gopikrishnan, P. Cizeau, M. Meyer, C.-K. Peng, and H. E. Stanley, Phys. Rev. E \textbf{60}, 1390 (1999).

\bibitem{gabaix03}
X. Gabaix, P. Gopikrishnan, V. Plerou, and H. E. Stanley, Nature \textbf{423}, 267 (2003).

\bibitem{ding83}
Z. Ding, C. W.  J. Granger, and R. F. Eagle, J. Empirical Finance \textbf{1}, 83 (1983).

\bibitem{ord85}
J. K. Ord, T. H. McInish, and R. A. Wood, J. Finance \textbf{40}, 723 (1985).

\bibitem{harris86}
L. Harris, J. Financ. Econ. \textbf{16}, 99 (1986).

\bibitem{schwert89}
G. W. Schwert, J. Finance \textbf{44}, 1115 (1989).

\bibitem{granger96}
C. W. J. Granger, and Z. Ding, J. Econometrics, \textbf{73}, 61 (1996).

\bibitem{plerou01}
V. Plerou, P. Gopikrishnan, X. Gabaix, L. A. Nunes Amaral, and H. E. Stanley, Quant. Finance \textbf{1}, 262 (2001).

\bibitem{plerou05}
V. Plerou, P. Gopokrishnan, and H. E. Stanley, Phys. Rev. E \textbf{71}, 046131 (2005).

\bibitem{bouchaud03}
J.-P. Bouchard, and M. Potters, \textit{Theory of Financial Risk and Derivative Pricing: From Statistical Physics to Risk Management} (Cambridge University Press, Cambridge, 2003).

\bibitem{black73}
F. Black, and M. Scholes, J. Polit. Econ. \textbf{81}, 637 (1973).

\bibitem{engle82}
R. F. Engle, Econometrica \textbf{50}, 987 (1982).

\bibitem{cox76}
J. C. Cox, and S. A. Ross, J. Finance Econ. \textbf{3}, 145 (1976).

\bibitem{campbell97}
J. Y. Campbell, A. W. Lo, and A. C. MacKinlay, \textit{The Econometrics of Financial Markets} (Princeton University Press, Princeton, 1997).

\bibitem{yamasaki05}
K. Yamasaki, L. Muchnik, S. Havlin, A. Bunde, and H. E. Stanley, Proc. Natl. Acad. Sci. U.S.A. \textbf{102}, 9424 (2005).

\bibitem{wang06}
F. Wang, K. Yamasaki, S. Havlin, and H. E. Stanley, Phys. Rev. E \textbf{73}, 026117 (2006);
F. Wang, P. Weber, K. Yamasaki, S. Havlin, and H. E. Stanley, Eur. Phys. J. B \textbf{55}, 123 (2007);
F. Wang, K. Yamasaki, S. Havlin, and H. E. Stanley, arxiv:0707.4638 (2007).

\bibitem{weber07}
P. Weber, F. Wang, I. Vodenska-Chitkushev, S. Havlin, and H. E. Stanley, Phys. Rev. E \textbf{76}, 016109 (2007).

\bibitem{kaizoji04b}
T. Kaizoji, Physica A \textbf{343}, 662 (2004).

\bibitem{kaizoji04}
T. Kaizoji, M. Kaizoji, Physica A \textbf{336}, 563 (2004).

\bibitem{dacorogna01}
M. M. Dacorogna, R. Gencay, U. A. Muller, R. B. Olsen and O. V. Pictet. \textit{An Introduction to High Frequency Finance} (Academic Press, London, 2001).

\bibitem{corral04}
A. Corral, Phys. Rev. Lett. \textbf{92}, 108501 (2004).

\bibitem{bunde2004}
A. Bunde, J. F. Eichner, S. Havlin, and J. W. Kantelhardt, Physica A \textbf{342}, 308 (2004).

\bibitem{bunde2005}
A. Bunde, J. F. Eichner, J. W. Kantelhardt, and S. Havlin, Phys. Rev. Lett. \textbf{94}, 048701 (2005).

\bibitem{livina2005}
V. N. Livina, S. Havlin, and A. Bunde, Phys. Rev. Lett. \textbf{95}, 208501 (2005).

\end{thebibliography}
\end{document}